# Efficient Personalization of Amplification in Hearing Aids via Multi-band Bayesian Machine Learning

*Aoxin Ni[1], Edward Lobarinas[2], and Nasser Kehtarnavaz[1]*

[1]Department of Electrical and Computer Engineering
[2]Callier Center for Communication Disorders
University of Texas at Dallas, Richardson, TX

*Abstract*— Personalization of the amplification function of hearing aids has been shown to be of benefit to hearing aid users in previous studies. Several machine learning-based personalization approaches have been introduced in the literature. This paper presents a machine learning personalization approach with the advantage of being efficient in its training based on paired comparisons which makes it practical and field deployable. The training efficiency of this approach is the result of treating frequency bands independent of one another and by simultaneously carrying out Bayesian machine learning in each band across all of the frequency bands. Simulation results indicate that this approach leads to an estimated hearing preference function close to the true hearing preference function in fewer number of paired comparisons relative to the previous machine learning approaches. In addition, a clinical experiment conducted on eight subjects with hearing impairment indicate that this training efficient personalization approach provides personalized gain settings which are on average six times more preferred over the standard prescriptive gain settings.

## I. Introduction

Widely used prescriptions of hearing aids, such as DSLv5 [1] and NAL-NL2 [2], involve setting gain values in a number of frequency bands based on a user's audiogram. An audiogram indicates the lowest level of sound pressure level (SPL) that a person can hear across audible frequency bands in a quiet audio environment [3]. There is a need to tailor the amplification function of hearing aids to noisy audio environments that are of particular interest to an individual user. Furthermore, with the recent introduction of more affordable Over-The-Counter (OTC) hearing aids, there is a growing need for their self-adjustment [4].

Any personalization or self-adjustment needs to be done in a simple and easy-to-use manner for it to be adopted by users. A complex personalization or self-adjustment involving too many "knobs" to adjust would be a major hindrance to its utilization. A number of simple methods, such as sliders, wheels, and pairwise comparisons, have been considered by researchers [5-8]. Among these methods, the pairwise comparison method is often chosen due to its simplicity in hearing preference studies, e.g. [9-12]. This method places minimal cognitive load on users as it merely involves selecting one out of two options similar to the pairwise comparisons in an eye exam.

A number of machine learning approaches have been developed in the literature to encode pairwise comparisons in a more systematic way and to conduct personalization of the amplification function of hearing aids including the ones by our research team [13-15]. These approaches normally involve a trade-off between the size of the search space and the duration of the fitting process or training. In our latest work, the machine learning approach of Maximum Likelihood Inverse Reinforcement Learning (MLIRL) [14, 15] was introduced in order to achieve personalization of amplification in an on-the-fly or online manner. Although this approach was shown to produce personalized settings that were preferred over the standard settings by about 10 times, it required a training duration of at least one hour to go through a large number of paired comparisons. This relatively long training time poses a bottleneck that restricts deployment in the field.

In this paper, a new machine learning approach is developed in order to address the above shortcoming. This approach involves conducting Bayesian machine learning in each frequency band in an independent manner. As a result, the number of paired comparisons to reach personalized settings is substantially reduced, thus making the online training process more feasible to deploy in the field.

The sections that follow are organized as follows. In section II, the independence aspect of the frequency bands is discussed. In section III, the developed Bayesian machine learning approach in each band is described allowing an efficient multi-band personalization. Then, the clinical setup for the subject testing conducted is mentioned in section IV. Finally, in section V, the results of our hearing preference and word recognition experiments are reported for eight subjects with hearing impairment. The paper concludes in section VI.

## II. Independence of Multi-Band Amplification

To personalize the amplification function in hearing aids, prescriptive gains across a number of frequency bands are taken to be the initial condition or the starting gain set. In this work, the DSLv5 hearing aid prescription is considered although any other prescriptions could be equally used to set the initial gain set. By adjusting these prescriptive gains within a lower and an upper boundary, a set of personalized gains or a personalized gain curve can be reached as shown in Figure 1. The lower boundary is established based on a user's audiogram. The upper boundary is determined by the loudness discomfort level (LDL) to ensure that the output audio signal does not exceed this level.

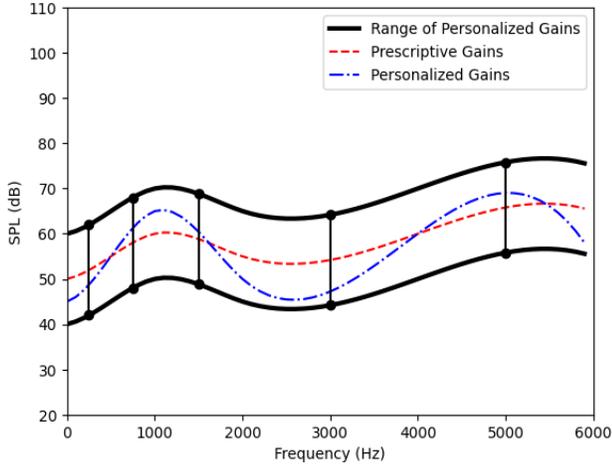

*Figure 1. Depiction of the range of personalized gains around the prescriptive gains.*

For the clinical experiment reported in this paper, the number of frequency bands is considered to be five corresponding to 0-500Hz, 500-1000Hz, 1000-2000Hz, 2000-4000Hz, and 4000-6000Hz. However, it should be noted that the developed personalization approach is general purpose in the sense that it can be applied to any number of frequency bands. Furthermore, the number of gain adjustment levels in each band is considered to be eight indicated by the set $\mathcal{X} = \{x_i \mid i = 1, ..., 8\}$ which correspond to these signal level changes {+12, +9, +6, +3, 0, -3, -6, -9} dB. Personalization of gains is expected to provide improved hearing in an audio environment in which a user is having difficulty hearing. Note that regardless of the number of gain adjustment levels, the developed personalization approach is applicable to any number of gain adjustment levels. The size of the space that needs to be searched to find the optimum setting is given by the number of levels raised to the power of the number of bands. For example, for five bands and eight adjustment levels, the size of the search space becomes $8^5$.

Personalization is achieved via paired comparisons of audio signals conducted by users. A paired comparison involves a user listening to an audio signal which is passed through two different gain sets. Then, the user selects the preferred gain setting from the pair. Paired comparisons are easy to conduct and demand a simple feedback from the user. For a search space of size $S$, it takes $S(S-1)/2$ paired comparisons to reach the optimum setting via the round robin tournament (RRT) method [10]. When the search space is too large, such as $8^5$ noted above, it would not be practical or feasible to expect users to go through a very large number of paired comparisons. In our previous MLIRL approach [14], only three levels were used to keep the search space size to $3^5$ or to keep the training time to about one hour for about 200 paired comparisons. One hour is a long time as far as users are concerned and thus a more efficient training approach is needed.

In order to reduce the number of paired comparisons, in this work, it is assumed that each band is independent of one another and via the simulations described next it is shown that this assumption of independence is valid. Let us consider an optimum gain set (personalized setting) that a user desires to reach. This set represents the gain values in the frequency bands generating a gain curve. Then, an RRT competition is conducted in the search space. Two tables are set up to keep track of two counts. Table I, named preference counting, is used to keep track of a count of the frequency band locations associated with the preferred hearing as determined by the cosine similarity of a gain curve with the optimum or true gain curve. Table II, named occurrence counting, is used to keep track of a count of the number of frequency band locations that the two curves differ.

TABLE I
EXAMPLE SHOWING COSINE SIMILARITY OF TWO HEARING PREFERENCE FUNCTIONS OR CURVES WITH TRUE GAIN CURVE

| True Gain Curve [3, 4, 5, 2, 3] | | | | |
|---|---|---|---|---|
| Curve 1 | Curve 2 | Similarity (True, 1) | Similarity (True, 2) | Preference |
| [4, **5**, **6**, 3, 4] | [4, **3**, **5**, **1**,4] | -1.23 | -1.03 | Curve 2 is preferred |

TABLE II
EXAMPLE OF PREFERENCE COUNTING TABLE

| Level Index | Band1 | Band2 | Band3 | Band4 | Band5 |
|---|---|---|---|---|---|
| 1 | 0 | 0 | 0 | 0 **+1** | 0 |
| 2 | 0 | 0 | 0 | 0 | 0 |
| 3 | 0 | 0 **+1** | 0 | 0 | 0 |
| 4 | 0 | 0 | 0 | 0 | 0 |
| 5 | 0 | 0 | 0 **+1** | 0 | 0 |
| 6 | 0 | 0 | 0 | 0 | 0 |
| 7 | 0 | 0 | 0 | 0 | 0 |
| 8 | 0 | 0 | 0 | 0 | 0 |

TABLE III
EXAMPLE OF OCCURRENCE COUNTING TABLE

| Level Index | Band1 | Band2 | Band3 | Band4 | Band5 |
|---|---|---|---|---|---|
| 1 | 0 | 0 | 0 | 0 **+1** | 0 |
| 2 | 0 | 0 | 0 | 0 | 0 |
| 3 | 0 | 0 **+1** | 0 | 0**+1** | 0 |
| 4 | 0 | 0 | 0 **+1** | 0 | 0 |
| 5 | 0 | 0**+1** | 0 **+1** | 0 | 0 |
| 6 | 0 | 0 | 0 | 0 | 0 |
| 7 | 0 | 0 | 0 | 0 | 0 |
| 8 | 0 | 0 | 0 | 0 | 0 |

TABLE III
EXAMPLE OF SIMULATED HEARING PREFERENCE FUNCTION

| Level Index | Band1 | Band2 | Band3 | Band4 | Band5 |
|---|---|---|---|---|---|
| 1 | 0.466 | 0.427 | 0.305 | 0.717 | 0.466 |
| 2 | 0.534 | 0.521 | 0.429 | **0.722** | 0.534 |
| 3 | **0.591** | 0.574 | 0.521 | 0.694 | **0.591** |
| 4 | 0.576 | **0.588** | 0.574 | 0.628 | 0.576 |
| 5 | 0.574 | 0.568 | **0.591** | 0.521 | 0.574 |
| 6 | 0.521 | 0.519 | 0.576 | 0.380 | 0.521 |
| 7 | 0.429 | 0.445 | 0.534 | 0.230 | 0.429 |
| 8 | 0.305 | 0.354 | 0.466 | 0.103 | 0.305 |

Table III illustrates an example of the simulation outcome. The above counting is repeated for all possible $8^5$ gain sets corresponding to the search space. Another table is then set up, based on the ratio of the two counts to represent the overall hearing preference curve for all possible gain sets. As shown in Table III, it can be seen that the frequency band locations of the overall hearing preference curve or gain set matches the frequency band locations of the true gain set. This simulation was repeated for a large number of different optimum or true gain sets and each time the same location matches were obtained, indicating that the frequency bands can be treated independently towards reaching the optimum or personalized gain set. It is worth mentioning that these simulations were repeated by using different adjustment levels and again the same location matches were obtained

### III. Personalization by Bayesian machine Learning

Considering the independence of frequency bands discussed above, the search space for finding the optimum gain set is drastically reduced for the paired comparisons needed just in a single band. For example, a total of only 8*7/2 = 28 paired comparisons would be needed for 8 levels.

Let $f(x)$ be a function indicating a user's hearing preference across gain adjustment levels in a frequency band which is to be determined by paired comparisons. Furthermore, let us consider that this function belongs to a Gaussian process $f(x) \sim \mathcal{GP}(0, k(x, x'))$ with the following kernel:

$$k(x, x') = \exp\left(-\frac{1}{2\lambda}(x - x')^2\right) \quad (1)$$

This kernel captures the distance between two gain settings where the parameter λ controls the preference function smoothness. Let the distribution of $f$ given $\mathcal{X}$ and λ be normal with zero mean as follows:

$$p(\mathbf{f} \mid \mathcal{X}, \lambda) = \mathcal{N}(\mathbf{f} \mid \mathbf{0}, \mathbf{K}) \quad (2)$$

where $\mathbf{f} = [f(x^1), \ldots, f(x^n)]$ represents a vector containing preference function values corresponding to $n$ adjustment levels, and the covariance $[\mathbf{K}]_{i,j} = k(x^i, x^j)$. Then, the optimal or personalized adjustment level is given by

$$x^* = \arg\max_x f(x) \quad (3)$$

Noting that a user's feedback is provided by paired comparisons, the so-called Probit likelihood for binary observation can be used to encode the user's feedback for the two options $x^a \in \mathcal{X}$ and $x^b \in \mathcal{X}$ in a paired comparison as described in [18]. The feedback is assigned -1 and +1 depending on which of the two gain sets is preferred by the user during a paired comparison episode $\mathcal{D} = \{d_k \in (-1,1) \mid k = 1, \ldots, m\}$, where $m$ denotes the number of paired comparisons in the episode. Now, according to [20], the probability that $x^a$ is preferred over $x^b$ can be written as follows:

$$p(d_k \mid \mathbf{f}_k, \sigma) = \Phi\left(d_k \frac{f(x_k^a) - f(x_k^b)}{\sqrt{2}\sigma}\right) \quad (4)$$

where Φ denotes the normal cumulative density function and the parameter σ denotes the degree of uncertainty or variations associated with a user's feedback. Hence, given the current preference function **f**, the likelihood of the feedback in an episode can be written as

$$p(\mathcal{D} \mid \mathbf{f}, \sigma) = \prod_{k=1}^m p(d_k \mid \mathbf{f}_k, \sigma) \quad (5)$$

Next the personalization problem is formulated within the Bayesian machine learning framework. That is the posterior probability of interest is obtained based on the prior probability and the feedback likelihood as follows:

$$p(\mathbf{f} \mid \mathcal{D}, \mathcal{X}, \lambda, \sigma) = \frac{p(\mathcal{D}|\mathbf{f},\sigma)p(\mathbf{f}|\mathcal{X},\lambda)}{p(\mathcal{D}|\mathcal{X},\lambda,\sigma)} \quad (6)$$

By using the Laplace Approximation discussed in [19], the posterior probability can be approximated as a normal density, that is

$$p(\mathbf{f} \mid \mathcal{D}, \mathcal{X}, \lambda, \sigma) \approx \mathcal{N}(\mathbf{f} \mid \hat{\mathbf{f}}, (\mathbf{K}^{-1} + \mathbf{W})^{-1}) \quad (7)$$

where **W** denotes the Hessian matrix. Then, an estimate of the function $\hat{\mathbf{f}}$ can be found in an iterative manner by using the Newton method via the following iterative equation shown in [19]:

$$\mathbf{f}^{new} = (\mathbf{K}^{-1} + \mathbf{W})^{-1}[\mathbf{W}\mathbf{f} + \nabla \log p(\mathcal{D} \mid \mathbf{f}, \sigma)] \quad (8)$$

As part of the above estimation, the parameters $\{\lambda, \sigma\}$ need to be determined. These parameters can be obtained via the maximum-a-posterior estimation method named L-BFGS-B described in [21, 22], that is

$$\{\lambda, \sigma\} \approx \arg\max_{\lambda,\sigma} p(\lambda, \sigma \mid \mathcal{D}, \mathcal{X}) \quad (9)$$

Table IV illustrates the Bayesian learning algorithm indicated above. The optimum or personalized adjustment levels are the ones corresponding to the highest hearing preference value in each band as illustrated in Figure 2.

TABLE IV
TRAINING OF PERSONALIZATION ALGORITHM

| |
|---|
| Input: Feedback from S(S-1)/2 pairwise comparisons |
| Initialize with the prescriptive gain set |
| For trajectory = 1: episodes |
|     Interact with audio environment S(S-1)/2 times |
|     Apply pairs and collect corresponding feedback |
|     For given parameter {λ,σ}, find an estimate $\hat{\mathbf{f}}$ using equation (8) |
|     Obtain updated parameters {λ,σ} using equation (9) |
| Output: $\mathbf{f}^*$ |

A simulation study was conducted to see whether the learning approach used generated an estimated preference function close to the true preference function. A vector of length 8 with random numbers between 0 and 1 was generated to represent the true preference function for 8 adjustment levels in a band. The 28 possible paired comparisons were divided into 7 episodes, each comprising 4 paired comparisons. Initializing with a flat preference function, the outcome of the simulation is shown in Figure 3. As can be seen from this figure, the estimated hearing preference function values **f** came close to the true hearing preference function values **f**.

If desired, it is possible to increase the accuracy of estimation by going through more episodes of paired comparisons. Based on a new episode of paired comparisons, the above Bayesian learning can be updated by beginning with the **f** values at the last episode. The parameters can also get computed based on the L-BFGS-B gradient for the new episode.

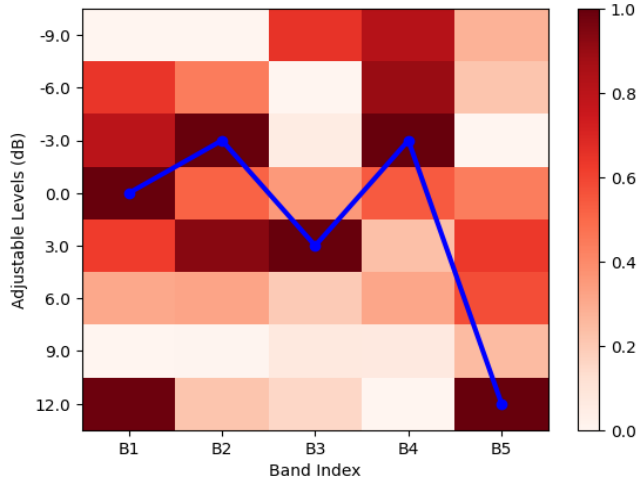

*Figure 2. Personalized gain curve obtained by connecting the highest hearing preference function value in each band*

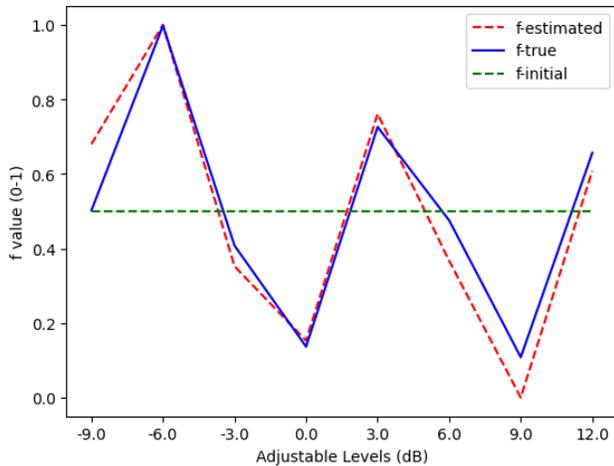

*Figure 3. Example of estimated hearing preference function in a band*

IV. CLINICAL EXPERIMENTAL SETUP

In addition to the simulations done above indicating the usability of the developed multi-band Bayesian machine learning approach, a clinical study has been conducted in this paper to show its applicability in practice. Eight subjects with mild to moderate hearing loss were recruited for this clinical study under an approved human subject Institutional Review Board protocol (IRB 20-13) at the University of Texas at Dallas. The eligibility for the participation included: (i) symmetric mild to moderate hearing loss, (ii) being able to speak and understand English, and (iii) being an adult in the age range of 21-80 years old who is capable of providing informed consent.

The clinical sessions were divided into the following three sessions: audiogram measurement session, training session, and testing session. The audiograms of the participants were obtained during the first session by a hearing healthcare professional. Based on the audiogram, a DSLv5 prescriptive gain set was generated from a web-based tool [23]. This DSLv5 gain set served as the baseline or the starting point of our personalization. The range of the personalization for each band was set to $(Gain_{prescription} - 9\ dB, Gain_{prescription} + 12\ dB)$ for the adjustable levels around the prescriptive gain set.

During the second session, the participants underwent a training procedure aimed at determining their individualized hearing preferences using the developed machine learning personalization. The participants wore a pair of commercial hearing aids in both ears linked via Bluetooth to a dedicated laptop placed in the sound booth in which the participants were sitting. Figure 4 provides an illustration of the experimental setup. The hearing aids were configured to deliver flat amplification through their internal processors with noise reduction and sound enhancement features disabled. The environmental microphones were also turned off so all audio was from the Bluetooth connection to the laptop. This setup ensured that any variations in hearing experience were solely attributable to different gain sets derived from the personalization algorithm. For guidance and communication, the experimenter remained stationed outside the sound booth and visible to the participant through a window. The experimenter ran the personalization algorithm running on a laptop (laptop 1 in Figure 4) which was connected to another laptop in the sound booth (laptop 2 in Figure 4) via an online meeting utility. Laptop 1 with the personalization algorithm controlled the amplification of audio signals through gain sets, transmitting audio signals from laptop 1 to laptop 2 via the online meeting utility, and then to the hearing aids via Bluetooth.

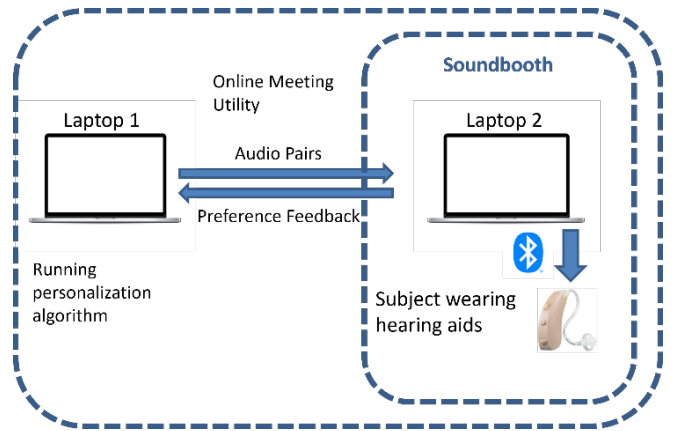

*Figure 4. Experimental setup of clinical testing*

The graphical-user-interface (GUI) of the personalization training program is shown in Figure 5. For each paired comparison, the participants were presented with a pair of audio signals generated by applying two different gain sets on the same audio signal of a sentence lasting approximately 2.5 seconds. Then, the participants were asked to indicate their hearing preference by selecting "Audio 1", "Audio 2", or "Same" in the interface. The audio signals corresponded to the TSP speech database which is widely used in speech processing [24]. This dataset consists of recordings of 1400 utterances spoken by 24 speakers. The recordings of this database are phonetically balanced and designed to contain a diverse range of speech sounds. Babble background noise was added to the spoken utterances with a moderate Signal-to-Noise Ratio (SNR) of 5 dB. During the training session, paired comparisons were conducted for a pair of gain sets. The gain values in the five bands indicated earlier were passed through a cosine curve interpolator to generate a smooth frequency response curve. Subsequently, a 64-tap FIR filter, obtained via a filter design module, was employed to generate the frequency response curve.

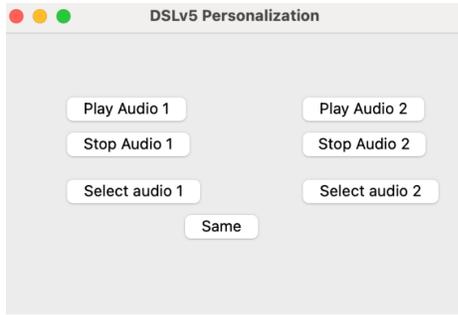

*Figure 3. Graphical-user-interface (GUI) of paired comparisons for obtaining the hearing preference function of a hearing aid user*

The training was carried out in 7 episodes with each episode corresponding to 4 paired comparisons. At the conclusion of each episode, the updated preference function as well as the updated parameters were used as the starting point for a subsequent episode. The preference function derived from the final episode was then used to establish the personalized gain settings. It is important to mention here that the entire training session was completed in less than 10 minutes, indicating the efficiency of the developed machine learning personalization algorithm.

## V. Subject Testing Results And Discussion

This section presents the outcomes of our clinical testing. Table 1 illustrates the audiograms of the participating subjects as well as the standard DSLv5 and personalized gain sets for the five frequency bands and eight adjustment levels considered. The results of the hearing preference testing, represented as percentages, are shown in Figure 6. As evident from this figure, all the participants indicated that the personalized setting was preferred over the standard, albeit with varying degrees depending on their level of hearing loss. On average across the eight participants, the personalized settings were favored six times more over the standard settings.

An additional experiment was conducted to assess whether the personalized settings had any adverse impact on word recognition or speech understanding. A 50-word list from the Northwestern University Auditory Test No. 6 (NU-6) dataset [25] was considered for this experiment. The list was played with babble background noise at 5dB SNR. Half of the words in the list were played using the standard DSLv5 setting, and the remaining half with the personalized setting. The order of presenting the words was randomized. The participants were instructed to repeat each word after it was presented to them once. Correct word repetition contributed to an increase in the word recognition score. The word recognition scores from this experiment are provided in Figure 7. As illustrated in this figure, the personalized setting exhibited no adverse impact on word recognition or speech understanding compared to the standard DSLv5 setting.

The work presented in this paper shows an efficient and effective approach to addressing the hearing healthcare needs of hard of hearing individuals by leveraging machine learning to personalize amplification. Our original hypothesis was that machine learning could generate an individualized prescription without degrading word recognition scores. By personalizing the hearing experience, users would likely be more satisfied with their devices and would be more likely to wear them. What our work demonstrates is that not only can machine learning derive and optimize individual preferences but that these preferences can also increase word recognition scores in competing background noise. The improvement in hearing in competing background noise addresses the top complaint among hearing aid users, difficulty in background noise. The approach presented in this paper addresses a major shortcoming of the current standard of care for hearing aid programming that remains tethered to average amplification targets and provides an efficient method that could substantially improve the current practice of hearing aid fitting.

TABLE V
STANDARD DSLV5 PRESCRIPTION AND PERSONALIZATION GAINS

| Subject | Audiogram (dB) | Standard Gains (dB) | Personalized Gains (dB) |
|---|---|---|---|
| 1 | 10, 15, 25, 45, 45 | 4, 2, 12, 30, 28 | 10, 5, 21, 21, 22 |
| 2 | 10, 25, 20, 30, 35 | 4, 9, 8, 22, 21 | 7, 18, 8, 28, 12 |
| 3 | 0, 5, 10, 30, 30 | -3, -5, 1, 22, 19 | -3, 7, 1, 22, 25 |
| 4 | 20, 20, 25, 25, 45 | 7, 6, 12, 21, 28 | 16, 18, 18, 21, 22 |
| 5 | 5, 0, 5, 30, 40 | 4, -4, -2, 22, 24 | 10, 5, 10, 25, 33 |
| 6 | 10, 15, 30, 40, 55 | 4, 2, 14, 27, 35 | 10, -7, 20, 21, 38 |
| 7 | 10, 10, 20, 15, 30 | 16, 0, 8, 13, 19 | 19, 12, -1, 22, 31 |
| 8 | 0, 30, 60, 40, 30 | 0, 11, 30, 27, 19 | -9, 14, 21, 30, 25 |

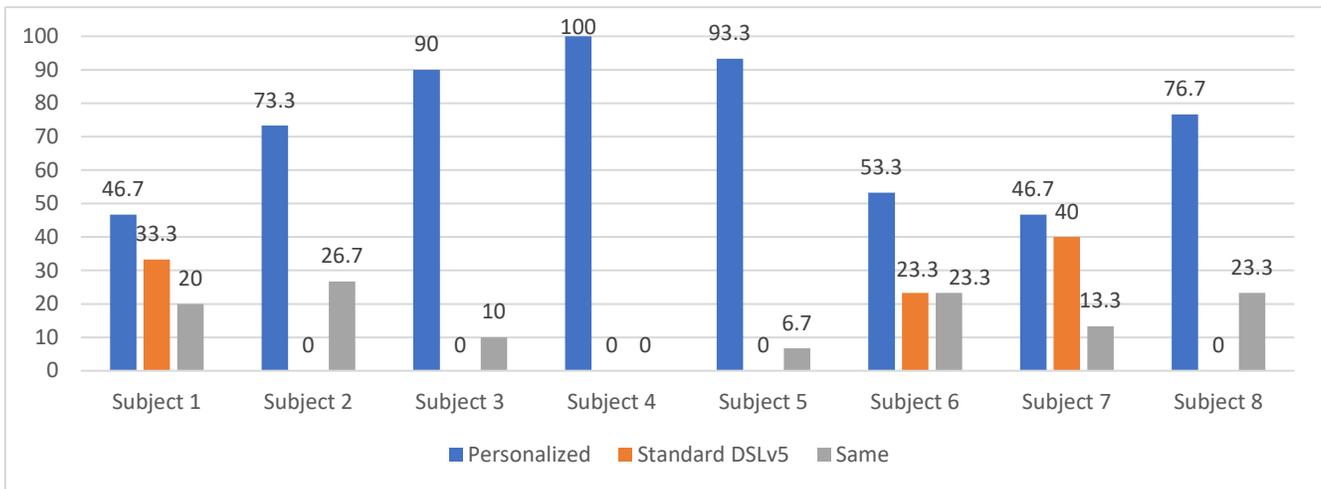

*Figure 6. Comparison of the personalized versus standard amplification settings; the numbers indicate the percentage of times the amplification setting was preferred during the testing session with Same indicating no preference.*

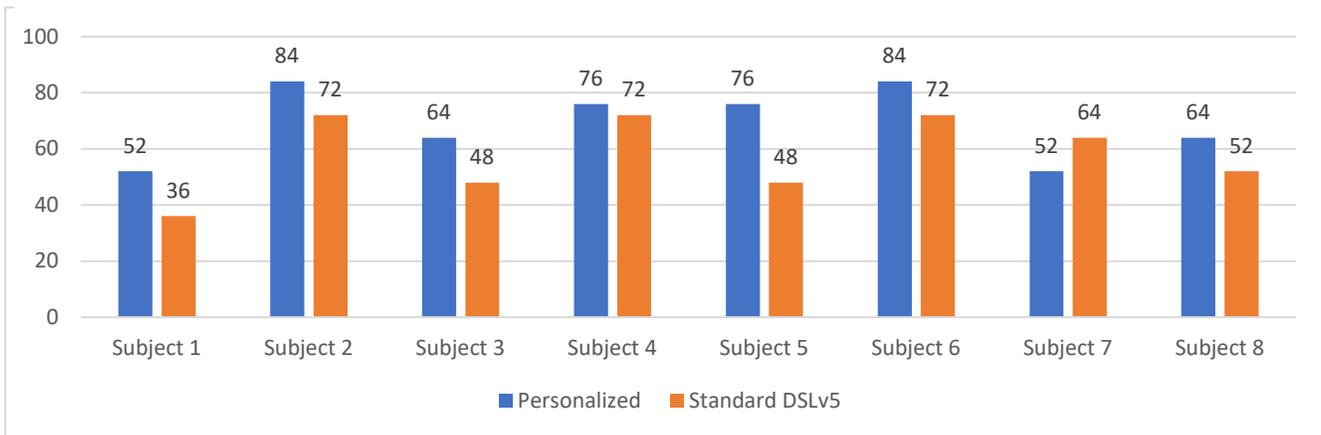

*Figure 7. Word recognition score percentages of the personalized and standard settings in babble background noise at 5dB SNR*

## VI. Conclusion

A training-efficient machine learning-based personalization approach has been introduced in this paper. This personalization approach involves the use of the Bayesian learning to model a hearing preference function corresponding to personalized gain values across any number of frequency bands that are considered to be independent of one another. The independence of the frequency bands has led to an efficient and thus practical training by conducting a small number of paired comparisons leading to reduced training times. The clinical experiment carried out on eight participants with hearing loss has shown that the personalized gain values are preferred over the standard DSLv5 prescriptive gain values on average by six times. In our future work, it is intended to turn this machine learning-based personalization approach into a smartphone app to enable its utilization in the field or real-world audio environments.

## Acknowledgment

The authors wish to thank Dr. Lindee Alvarez for her help in conducting the clinical experiments.